\documentclass[10pt,a4paper,twocolumn]{scrartcl}
\usepackage[T1]{fontenc}
\usepackage[latin1,utf8]{inputenc}
\usepackage{lmodern}
\usepackage{enumitem}
\usepackage[pdftex]{graphicx}
\usepackage[ngerman,spanish,english]{babel}
\usepackage[tracking=true]{microtype} % improves typesetting                        
\usepackage{amsmath,amssymb,amsthm}
\usepackage{mathrsfs}
\usepackage{mathtools}
\usepackage{grffile}   	% picture names with special types
\usepackage{bbm} 		%for use of identity-matrix
\usepackage{dsfont}
\usepackage[export]{adjustbox}[2011/08/13]
\usepackage[]{subfigure}
\usepackage{verbatim} 
\usepackage{color}
\usepackage{hyperref}
\usepackage{accents}
\usepackage{textcomp}
\usepackage{multirow}
\usepackage{booktabs}
\usepackage{geometry}
\usepackage{cite}
\usepackage{wrapfig}
\usepackage{appendix}
\usepackage[framed,numbered,autolinebreaks,useliterate]{mcode} %Paquete para código de MATLAB.
\usepackage{lipsum, babel}
\usepackage{calligra}
\DeclareMathAlphabet{\mathcalligra}{T1}{calligra}{m}{n}
\DeclareFontShape{T1}{calligra}{m}{n}{<->s*[2.2]callig15}{}
\usepackage[autolinebreaks]{mcode}
\newcommand{\sff}[1]{\ensuremath{\mathcalligra{#1}}}

\usepackage{etoolbox}
\begin{document}

%Macros are not universal, but many of the following macros have been used for more than 20 years by YO and quite useful.
%Space/Indentation
\newcommand{\noi}{\noindent}
\newcommand{\vnu}{\vspace*{-1mm}}
\newcommand{\vnd}{\vspace*{-2mm}}
\newcommand{\vnt}{\vspace*{-3mm}}
\newcommand{\vnq}{\vspace*{-4mm}}
\newcommand{\vnc}{\vspace*{-5mm}}
\newcommand{\vu}{\vspace*{1mm}}
\newcommand{\vd}{\vspace*{2mm}}
\newcommand{\vt}{\vspace*{3mm}}
\newcommand{\vc}{\vspace*{5mm}}
\newcommand{\hs}{\hspace*{5mm}}

%Equations/Entries
\newcommand{\beq}{\begin{equation}}
\newcommand{\bequo}{\begin{quotation}}
\newcommand{\beqa}{\begin{eqnarray}}
\newcommand{\eeq}{\end{equation}}
\newcommand{\equo}{\end{quotation}}
\newcommand{\eeqa}{\end{eqnarray}}
\newcommand{\non}{\nonumber}
\newcommand{\mx}{\mbox}
\newcommand{\mxf}[1]{\mbox{\footnotesize{#1}}}
\newcommand{\lb}{\label}
\newcommand{\fr}[1]{(\ref{#1})}
\newtheorem{entry}{}[section]
\newcommand{\bent}[1]{\vspace*{-2cm}\hspace*{-1cm}\begin{entry}\lb{e{#1}}\rm}
\newcommand{\eent}{\end{entry}}
\newcommand{\fre}[1]{{\bf\ref{e{#1}}}}
\newcommand{\Emark}{$\sqcap\hspace{-2.7mm}\sqcup$}

%Greek Letters
\renewcommand{\a}{\alpha}
\renewcommand{\b}{\beta}
\newcommand{\g}{\gamma}
\newcommand{\G}{\Gamma}
\renewcommand{\d}{\delta}
\renewcommand{\th}{\theta}
\newcommand{\Th}{\Theta}
\newcommand{\D}{\Delta}
\newcommand{\e}{\epsilon}
\newcommand{\s}{\sigma}
\renewcommand{\S}{\Sigma}
\newcommand{\w}{\omega}
\newcommand{\W}{\Omega}
\newcommand{\al}{\alpha}
\newcommand{\bet}{\beta}
\newcommand{\gam}{\gamma}
\newcommand{\lam}{\lambda}
\newcommand{\Lam}{\Lambda}
\newcommand{\eps}{\epsilon}
\newcommand{\cmark}{\ding{51}}%
\newcommand{\xmark}{\ding{55}}%
\newcommand*\cc[1]{\omit\hfil$\displaystyle#1$\hfil}

%Letter Style
\renewcommand{\vec}{\bm}
\newcommand{\svec}[1]{\mbox{{\footnotesize $\bm{#1}$}}}
\newcommand{\bm}[1]{\mbox{\boldmath $#1$}}
\newcommand{\sbm}[1]{\small{\mbox{\boldmath $#1$}}}
\newcommand{\fns}{\footnotesize}
\newcommand{\ol}{\overline}
\newcommand{\ul}{\underline}
\newcommand{\vr}{\bm{r}}
\newcommand{\vq}{\bm{q}}
\newcommand{\vk}{\bm{k}}
\newcommand{\vx}{\bm{x}}
\newcommand{\vy}{\bm{y}}
\newcommand{\ff}[1]{\mathbb{#1}}
 
%Arrows
\newcommand{\see}{$\rightarrow$}
\newcommand{\react}{$\longrightarrow$}
\newcommand{\map}{\rightarrow}
\newcommand{\pam}{\leftarrow}
\newcommand{\maps}{\rightarrow}
\newcommand{\imply}{\Rightarrow}
\newcommand{\Lrar}{\Leftrightarrow}
\newcommand{\rar}{\rightarrow}
\newcommand{\irrev}{\prec\prec}
\newcommand{\go}{\prec}
\newcommand{\adeq}{\stackrel{A}\sim}
\newcommand{\theq}{\stackrel{T}\sim}

%Derivatives
\newcommand{\pder}[2]{\frac{\partial {#1}}{\partial {#2}}}
\newcommand{\pdert}[2]{\frac{\partial^2 {#1}}{\partial {#2}^2}}
\newcommand{\fder}[2]{\frac{\delta {#1}}{\delta {#2}}}
\newcommand{\PDD}[3]{\left.\frac{\partial^{2}{#1}}{\partial{#2}^{2}}\right|_{#3}}
\newcommand{\PD}[3]{\left.\frac{\partial{#1}}{\partial{#2}}\right|_{#3}}
\newcommand{\der}[2]{\frac{d {#1}}{d {#2}}}

%Parentheses
\newcommand{\av}[1]{\left\langle{#1}\right\rangle}
\newcommand{\la}{\langle}
\newcommand{\ra}{\rangle}
\newcommand{\bra}[1]{\langle{#1}|}
\newcommand{\ket}[1]{|{#1}\rangle}
\newcommand{\FB}{${\bm \la\hspace{-.7mm}\la}$}  %footnote title start
\newcommand{\FE}{${\bm \ra\hspace{-.7mm}\ra}$}
\renewcommand{\v}{\vspace*{2mm}}
%logic
%\renewcommand{\or}{\vee} %this interferes with beqa/eeqa
\renewcommand{\and}{\wedge}

\renewcommand{\deg}{^\circ}
\newcommand{\com}{{\bf [C] }}
\newcommand{\cend}{\Emark\[\]\vspace*{-1. cm}}
\newcommand{\x}{\times}
\newcommand*{\bfrac}[2]{\genfrac{}{}{0pt}{}{#1}{#2}}

\newcommand{\myfrac}[3][0pt]{\genfrac{}{}{}{}{\raisebox{0pt}{$#2$}}{\raisebox{-#1}{$#3$}}}

%%%%%%%%

\hypersetup{
pdftitle={The distortion tensor in cubic slip-systems},
pdfauthor={Danyel Cavazos-Cavazos and Flavio F. Contreras-Torres},
pdfsubject={Contrast factors in ionic systems}
}

\twocolumn[\begin{@twocolumnfalse}

%Analytical form for the dislocation field in cubic slip-systems: application to the microstructural characterization of rock-salt (Fm$\bar{3}$m}) materials

\section*{\Large{Determination of contrast factors for cubic slip-systems and their application in the microstructural characterization of binary Fm$\mathbf{\bar{3}}$m} materials}      	% make a section
 
% if you what a title page search for \maketilte

 \vspace{3mm}  	%adds a vertical space

\normalsize 		%sets size of writing

\large{Danyel Cavazos-Cavazos$^1$ and Flavio F. Contreras-Torres$^{1,2,}$*} \newline

$^1$Centro del Agua para América Latina y el Caribe, Tecnológico de Monterrey, Monterrey 64849, Mexico\newline

$^2$Departamento de Física, Tecnológico de Monterrey, Monterrey 64849, Mexico\newline

* corresponding author: contreras.flavio@itesm.mx

\vspace{0mm}

\section*{Abstract} % the * in \section*{Abstract:} makes that this section will not be listed if you do a \tableofcontents.

The stress-strain anisotropy ---including both elastic properties and dislocation distributions--- of crystalline systems can be studied using a dislocation model proposed by Ungár and Borbély. However this model requires fundamental parameters (i.e., contrast factors), which take into account both the elastic properties and symmetry for each system. In this study, such contrast factors were calculated from first principles for cubic slip-systems through the evaluation of the distortion tensor in a dislocation-dependent coordinate set. Moreover, a straightforward expression for the computation of contrast factors is provided. Further analysis of the microstructural parameters was carried out using the modified Williamson-Hall method to characterize rock-salt materials.

\vspace{7mm}

\begin{tabular}{p{1,5cm}p{12,55cm}p{1,5cm}}   %begins tabular

% to center the Keywords i did a table with 3 columns and used the first and the last one as free space

& \textbf{Key Words}: X-ray line profile analysis $\cdot$ Strain anisotropy $\cdot$ Dislocation field $\cdot$ Contrast factor $\cdot$ Williamson-Hall & \\ & \\
\midrule
\end{tabular}  % ends tabular

\vspace{7mm}

\end{@twocolumnfalse}]

\section*{Introduction}

Diffraction methods are commonly used to characterize the microstructure of materials. X-ray diffractometry, for example, is a powerful tool to study the shape, size and distribution of crystallites; lattice faults and twinings; and the arrangement and density of stress-strain dislocations \cite{1,12,13A}. All the above information is simultaneously embedded within the sample's diffractogram, and thus several approaches to estimate apparent size parameters and mean square strain values have been proposed along the last few decades. The Williamson-Hall (WH) and the Warren-Averbach (WA) are two classical methods \cite{3C,3B,3A,4,3D} that can describe the microstructure for bulk materials. However, several assessments to obtain microstructural information resulted in the lack of a monotonic behavior as evidenced mainly for WH \cite{X1,X2,X3}. To effectively decouple the sample's size and strain contributions, Ungár and Borbély \cite{5A,10} modified the WH and WA methods making use of contrast factors ($C_{hkl}$) which account for the stress-strain anisotropy introducing scaling factors. As a result, the accuracy and agreement of the microstructural parameters estimated by diffraction techniques is highly improved, and even can be compared to those measurements carried out by transmission electron microscopy \cite{5B}.

A fundamental step to implement the proposed gauge relies on the evaluation of the distortion tensor in an anisotropic medium. By making use of the Lekhnitskii complex potential \cite{Td,11E}, the Stroh dislocation eigenvalues\cite{5H} and a dislocation-dependent coordinate system, a straightforward expression for the computation of this parameter can be obtained. Thus, the elastic component of dislocation contrast factors can be evaluated in the system's proper coordinates. In particular, the rock-salt (Fm$\bar{3}m$) structures impose additional restrictions on their dislocation allowance mainly due to the crystal’s overall neutral charge. Even though a parametric evaluation of contrast factors on cubic symmetries has been previously implemented\cite{5D}, its application does not transition directly into materials with more limiting degrees of freedom. Therefore the use of a first principles approach to calculate contrast factors for such materials is necessary, with different constraints arising for each symmetry/slip-system coupling.

Herein, the distortion tensor for cubic slip-systems is described using a stretched coordinate system that allowed to compute the elastic component of the contrast factors on each dislocation in a more simple way. In particular, the contrast factors for binary Fm$\bar{3}$m systems are calculated from first principles and used to characterize the microstructure of KCl and NaCl materials through a modified WH analysis.

\begin{comment}
The aim of this study was to evaluate the microstructure of materials with a symmetry outside the  regular applicability rage (in particular, rock-salt structured materials) through  advanced, X-ray line profile analysis techniques. The additional symmetry constrains for such materials required a first principles calculation of contrast factors.     
\end{comment}
\section*{Theoretical basis}

\subsection*{Modified WH method}

The classic WH method\cite{3D} resolves both size ($\beta_{p}$) and strain ($\beta_{s}$) broadening contributions on real crystals by taking advantage of their different order dependence with respect to Bragg's angle ($\theta$). The former contribution occurs due to the finite size effects of the diffracting system, while the later one arises from its lattice distortions. %Cite procedings, libro aqua% 
More specifically,

\begin{equation}
\label{WH}
\tilde{\beta} = \frac{1}{\tau} + 2\,\zeta\,\tilde{d}
\end{equation}

\noi where $\tilde{\beta} = \beta cos(\theta) / \lambda$ and $\tilde{d} = 2sin(\theta)/\lambda$ are respectively the sample-related broadening and plane spacing, described in a reciprocal space and $\lambda$ is the source wavelength; $\tau$ is the \textit{apparent} crystallite size as originally defined by Jones \cite{3C} while $\zeta$ is the \textit{apparent} strain as originally defined by Stokes and Wilson\cite{3B}. Both $\tau$ and $\zeta$ are the integral breaths of the sample's crystallite size and the tensile strain distributions, respectively. In particular, $\tau$ was interpreted by Hall\cite{3A} as a characteristic length scale for the lattice regions which diffract coherently within the system, and $\zeta$ depends directly on the distribution curve which governs the system's strain. If the distribution is for example uniform and isotropic, $\zeta = 2\epsilon$, where $\epsilon$ is the maximum relative displacement $(\Delta d/d\equiv \Delta \tilde{d}/\tilde{d})$ of a lattice point from its ideal position.    
The modified WH method \cite{5A} broadens the scope of the original approach into systems on which strain anisotropy is significant by proposing a proper scaling factor $\delta = \tilde{d}\sqrt{C}$ instead of $\tilde{d}$ as in eq.\eqref{WH}, with $C$ being the average dislocation contrast factor. More significantly, $\tilde{\beta}(\delta)$ is to take a quadratic form as in
\begin{equation}
\label{MWH}
\tilde{\beta}(\delta) = \tilde{\beta_0}+\tilde{\beta_1}\delta + \tilde{\beta_2}\delta^2
\end{equation}

\noi where $\tilde{\beta_0}\equiv 1/\tau'$, $\tilde{\beta_1}\equiv 2\zeta'$ and $\tilde{\beta_2}\propto \sqrt{\strut Q}$, where $\tau'$ and $\zeta'$ are the modified WH parameters and $Q$ is the the correlation coefficient between adjacent lattice points, often interpreted as the fluctuation $Q = \av{\rho^2} - \av{\rho}^2$ of the dislocation density, $\rho$, \cite{11F}.
In particular, eq. \eqref{MWH} reduces to a linear case as in eq. \eqref{WH} when  $Q$ is zero or negligible.

\subsection*{Contrast Factors}

The computation of the average contrast factor $C$ from each individual contrast factor $C_{hkl}$, as extensively described by Armstrong and Lynch\cite{11A}, can be performed as:
\begin{eqnarray}
\label{C}
C\equiv \av{C_{hkl}} &=& \frac{1}{N} \sum\limits_{i=1}^N\; C^i_{hkl} \nonumber \\
&=&\frac{1}{N} \sum\limits_{K,L=1}^6\sum\limits \; G^i_{KL} E^i_{KL}
\end{eqnarray}
where $K$,$L$ are the indices for the reduced form of each 4-rank tensor and $N$ is the total number of degenerate slip systems (see Appendix 1). If not all slip-systems are equally populated, appropriate weight factors for each system should be calculated and included to the overall ensemble %\eqref{C} 
\cite{11A}. Specifically, the RHS of eq. \eqref{C} is split into a geometric component $\vec{G}\equiv G_{KL} = G_{ijkl}$ and an elastic one $\vec{E}\equiv E_{KL} = E_{ijkl}$, respectively described by:

\begin{equation}
\label{G}
G_{ijkl} = \gamma_{i}\,\gamma_{j}\,\gamma_{k}\,\gamma_{l}
\end{equation}
and
\begin{equation}
\label{E}
E_{ijkl} = \frac{1}{\pi} \int\limits_0^{2\pi} \; T_{ij}\,T_{kl}\, d\varphi
\end{equation}

\noi with $\gamma_i$ being the direction cosine between  the scattering vector and the slip coordinate system for a particular geometry and dislocation type (see \cite{11A}) and $T_{ij}$ is the distortion tensor associated with the displacement field of each dislocation. 

The calculation of the geometric components for each contrast factor is relatively straightforward\cite{5H}
and thus it is the evaluation of eq. \eqref{E} which turns out computationally demanding. Even though most calculations of contrast factors as pioneered by Borbely et.al. make use of numerical methods to perform this task\cite{5C}, an analytical form for the distortion tensor can be obtained by introducing a 'stretched' coordinate system which readily integrates information about a particular dislocation ---this within the Stroh-Lekhnitskii formalism. Moreover, $T_{ij}$ can be written as a linear combination of the distortion displacements, which take a closed, symmetric form when expressed in this 'stretched' coordinates as will be described next.

\subsection*{Displacement Field}

In order to evaluate eq. \eqref{E}, both the system's distortion tensor $T_{ij}$ within the dislocation plane and the overall 3D displacement field $u_\kappa$ must be evaluated. The later has been previously described by Teodosiu, with an explicit form given by \cite{Td}:
\begin{equation}
u_\kappa (x_1,x_2) = \frac{1}{\pi} \, \ff{I}\text{m} \left\{ \sum\limits_{\a=1}^3 A_{\kappa,\a} \; D_\a \; F_\a \right\}
\end{equation}

\noi with $F_\a = \ln(x_1 + p_\a x_2)$ being the Lekhnitskii complex potential and $p_\a$ being the correspondent root of the sextic polynomial associated with each slip system, while $A_{\kappa,\a}$ and $D_\a$ are the eigenvalues to the corresponding Stroh eigenvalue equation associated with each dislocation.%Cite Scardi%
By realizing the complex nature of $A$, $D$, and $F$ (i.e. their real and complex components), an explicit form for the displacement field can be written as:

\begin{eqnarray}
u_\kappa  &=& \frac{1}{\pi} \,  \sum\limits_{\a=1}^3 \left\{ A_{\kappa,I}\,D_R\,F_R + A_{\kappa,R}\,D_I\,F_R \right. \nonumber \\
&\,& \left.\;\;\;\;\;+ A_{\kappa,R}\,D_R\,F_I - A_{\kappa,I}\,D_I\,F_I\right\}
\end{eqnarray}

\noi where the $X_R$ and $X_I$ subscripts denote real and imaginary part, respectively. The $\a$ subscript  for each term of the equations above has been dropped for simplicity.

\section*{Distortion Tensor in a stretched coordinate system}

\begin{figure}[t!]   
% includes a figure in your pdf, the pictures should be in the image folder of your project.
	\includegraphics[width=0.45 \textwidth]{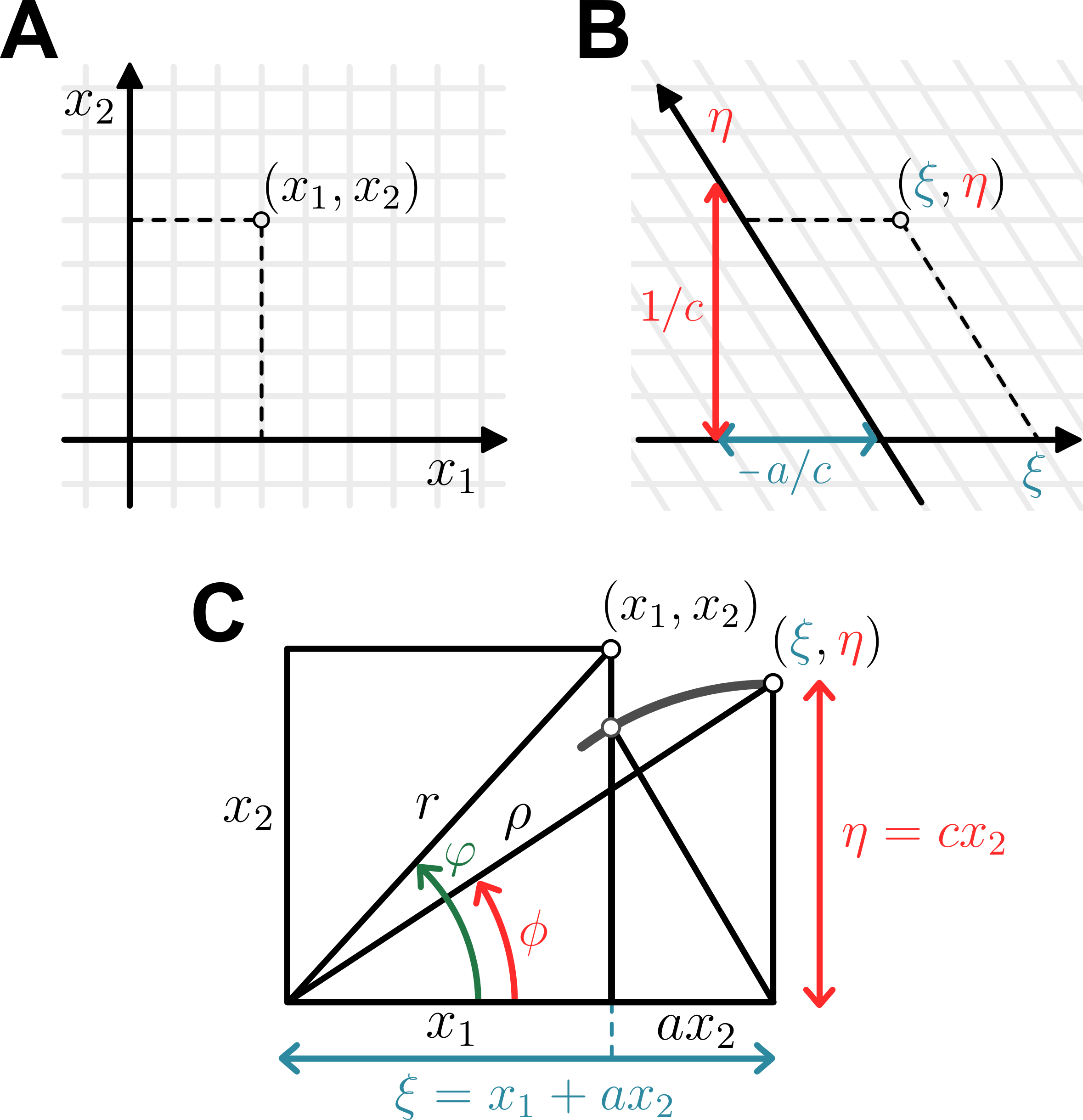}
	\caption{Cartesian (A) and 'Stretched' (B) coordinate axes system for a particular Stroh dislocation eigenvalue $p_\a=a+ic$. (C) Geometrical relationships between both coordinate systems.}
	\label{axes}  % give your figure a lable to cite it with \ref{Fig:beach}
\end{figure}

An analytical form for the dislocation tensor can be obtained introducing a \textit{stretched} coordinate system $(x_1,x_2), (r,\varphi) \leftrightarrow (\xi,\eta), (\rho,\phi)$ based upon each Stroh dislocation eigenvalue $p_\alpha = a+ic$ (see Fig. \ref{axes}), with: 

\begin{align}
\label{stretched}
  \begin{cases}
      \xi = x_1 + ax_2 & \rho^2 = \xi^2+\eta^2  \\
      \eta = cx_2 & \phi = \arctan(\eta/\xi)  
    \end{cases}
\end{align}

\noi and the corresponding transformation relation:

\begin{equation}
\begin{pmatrix} \xi \\\eta \end{pmatrix} = \begin{pmatrix} 1 & a  \\0&c \end{pmatrix}\cdot\begin{pmatrix} x_1 \\x_2 \end{pmatrix}
\end{equation}

\noi In this new coordinate system, the real and imaginary parts for the distortion displacement $H_{i} = r \partial_iF$ with $\partial_i F_{} = \pder{F_{}}{x_i}$ and $\sff{r} = r/\rho^2$
%= \pder{F_{}}{\gamma}\,\pder{\gamma}{x_i} $ and $\sff{r} =r/\gamma$ 
can be succinctly written as:
\begin{eqnarray}
\label{H}
 H_{1,R}  = \phantom{-}\xi\,\sff{r} &\;\; & H_{2,R}  = (a \xi + c \eta)\,\sff{r}   \nonumber \\
H_{1,I}  = -\eta\,\sff{r} &\;\;& \, H_{2,I}  = (c \xi - a \eta)\,\sff{r} 
\end{eqnarray}

\noi Again, the $\a$ subscripts  were dropped  for simplicity. Under the previous definitions, the close-form equation for  the distortion tensor is given by:
\begin{eqnarray}
\label{T}
T_{ij}  &=& \frac{2\pi\,r}{b} \partial_j u_i \nonumber \\
&=&  \,  \frac{2}{b}\sum\limits_{\a=1}^3 \left\{ A_{i,I}\,D_R\,H_{j,R} + A_{i,R}\,D_I\,H_{j,R} \right. \nonumber \\
&\,& \left.\;\;\;+ A_{i,R}\,D_R\,H_{j,I} - A_{i,I}\,D_I\,H_{j,I}\right\}
\end{eqnarray}

It can be shown (see Appendix 2) that the distortion displacement (and thus the distortion tensor) can be expressed as a function of the polar angle $\varphi$ only. Thus, eq. \eqref{E} can equivalently defined in terms of the 'stretched' polar angle as:

\begin{equation}
\label{E2}
E_{ijkl} =  \frac{1}{\pi} \int\limits_0^{2\pi} \; T_{ij}\,T_{kl}\, d\phi
\end{equation}

Finally, the eigenvalues $A_{\kappa,\alpha}$ and $D_{\alpha}$ are quantities which directly depend on the sample's elastic constants, and they can be directly calculated from the reduced elastic compliances and $p_\alpha$\cite{11A}. 

\begin{comment}
\subsection*{Slip systems in ionic (Fm\texorpdfstring{\textbf{$\bar{3}$}}{TEXT}m) structures}
\end{comment}
  
\section*{Application to binary (Fm\texorpdfstring{\textbf{$\mathbf{\bar{3}}$}}{TEXT}m) structures}
\subsection*{Materials and methods}
Crystalline samples of KCl and NaCl were prepared by hand-milling in an agathe mortar.  To remove any adsorbed water a subsequent heat treatment was performed at 450°C during 48h. For diffraction measurements the powders were deposited on a silicon plate. A ready-made plaque of crystalline VO was also measured for comparison.

X-ray diffraction measurements were performed on a Bruker X-ray D2 Phaser powder diffractometer using $\beta$-filtered, Cu-$K_\alpha$ radiation ($\lambda$ = 0.154 nm; 30kV, 10mA). Data were recorded with a step size of 0.010° and an integration time of 1 s from 20° to 100° (2$\theta$).  The instrumental broadening was estimated with a corundum (Al$_2$O$_3$) standard with measured instrumental function parameters of $u=8.2\times10^{-2}$, $v=-3\times10^{-4}$ and $w=8.0\times10^{-6}$. The instrumental contribution was subtracted from the raw data before further analysis. The integral breaths estimations were performed with the EVA Diffract.Suite\texttrademark\; and the computation of contrast factors was performed with code written in MATLAB.

\subsection*{Results and Discussion}

\begin{figure}[t!]   
% includes a figure in your pdf, the pictures should be in the image folder of your project.
	\includegraphics[width=0.45 \textwidth]{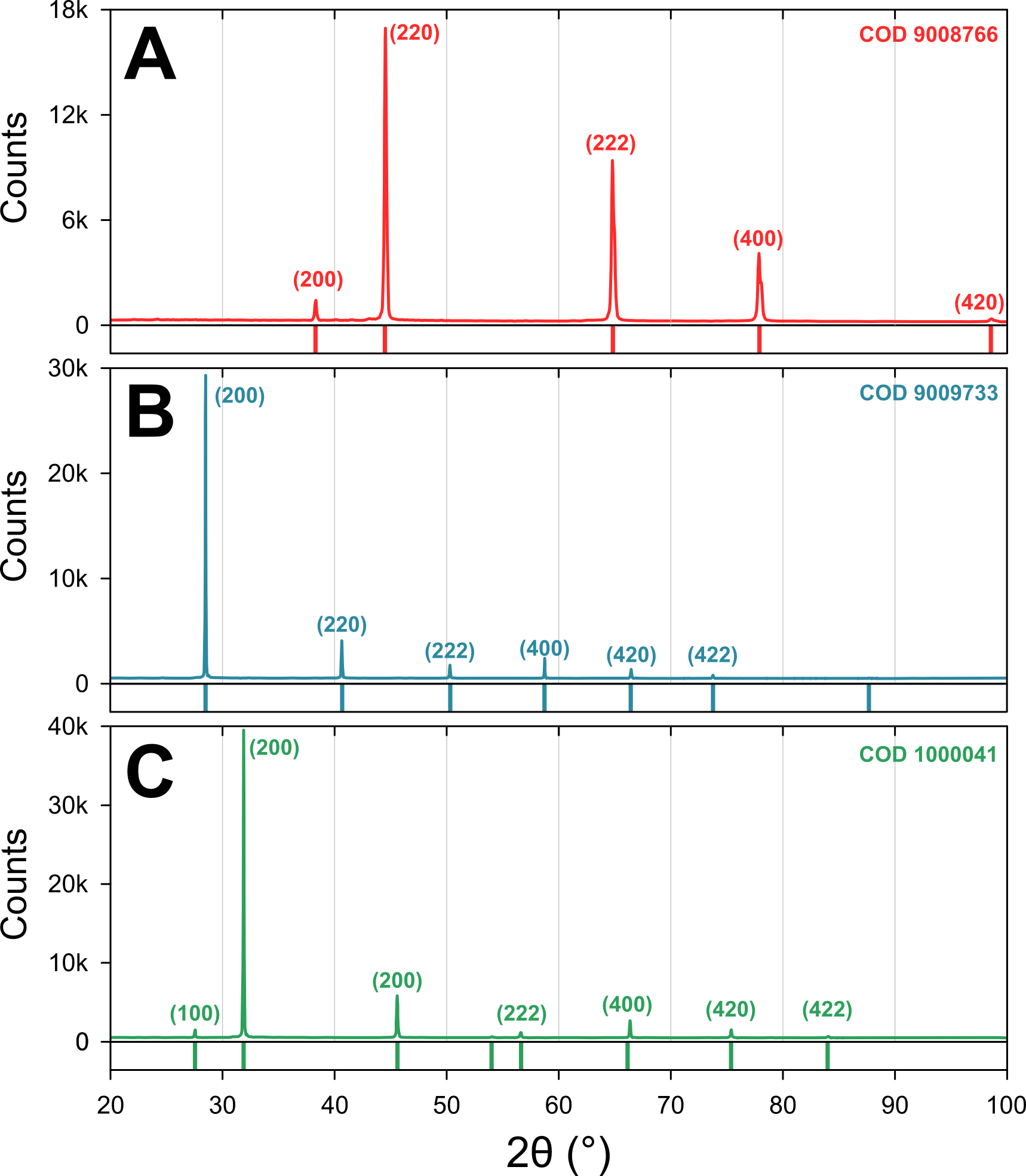}
	\caption{Recorded diffractograms for (A) a VO plaque,  and powder samples of (B) KCl and (C) NaCl. Their respective COD card is included at the bottom of each plot for reference.}
	\label{diffs}  % give your figure a lable to cite it with \ref{Fig:beach}
\end{figure}

The recorded X-ray measurements for both the VO plaque and the powder samples are shown in Fig. \ref{diffs}. The diffractograms showed a high degree of crystallinity and a match to their respective COD information card\cite{CD}. The calculated lattice constants were 0.45 nm, 0.63 nm an 0.56 nm for VO, KCl and NaCl, respectively, in good agreement with the compound's reported lattice parameters.

\begin{figure*}[t!]   
% includes a figure in your pdf, the pictures should be in the image folder of your project.
	\includegraphics[width=1 \textwidth]{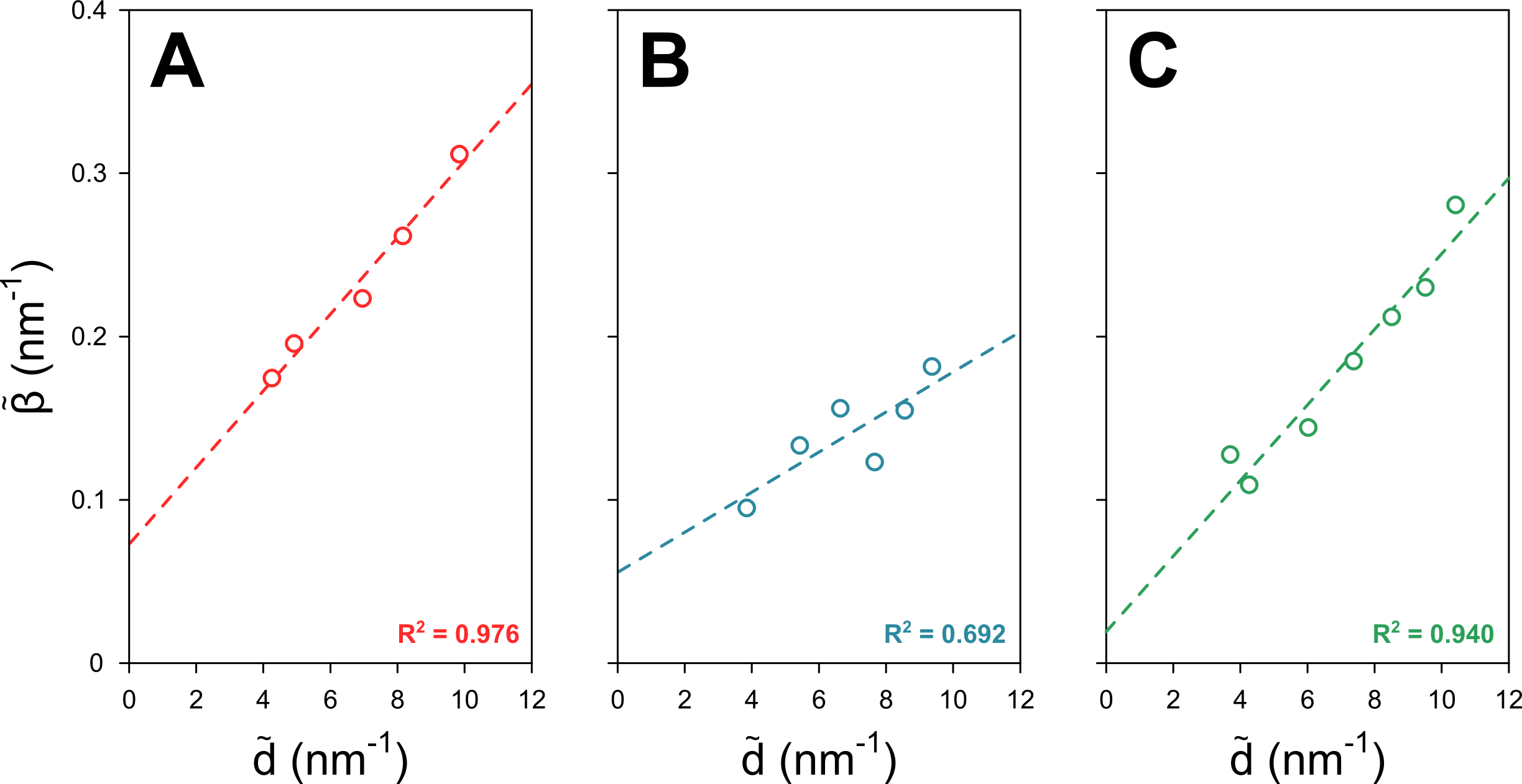}
	\caption{Classical WH plots in reciprocal space for (A) a VO plaque,  and powder samples of (B) KCl and (C) NaCl.}
	\label{WHp}  % give your figure a lable to cite it with \ref{Fig:beach}
\end{figure*}

The measured integral breaths for all three samples are presented in Fig. \ref{WHp}
in the form of classical WH plots. On the one hand, the VO plaque showed a relatively linear behavior, in good agreement with the behavior described by the classical WH analysis (as in eq. \ref{WH}). The ionic powder samples, on the other hand, seem to show a lesser monotonic behavior as evidenced in a lower coefficient of determination, particurlarly for KCl, which has been previously associated with strain anisotropy \cite{10}. Therefore, a modified WH analysis for both samples was implemented.

\begin{table}[b!]
\setlength{\tabcolsep}{8pt}
\renewcommand{\arraystretch}{1.3}

\centering
  \caption{Average Contrast Factors $C\equiv\av{C_{hkl}}$ for both KCl and NaCl$^a$}
    \begin{tabular}{ccccc}
    \toprule
    %\multicolumn{7}{c}{Average Contrast Factors $\av{C_{hkl}}$} \\
    %\midrule
    & \multicolumn{2}{c}{KCl} & \multicolumn{2}{c}{NaCl}\\
    $(hkl)$ & Edge & Screw  & Edge & Screw  \\
    \midrule
    100 & 0.1857  & 0.1007 & 0.1870  & 0.1334    \\
    200 & 0.1857  & 0.1007 & 0.1870  & 0.1334   \\
    220 & 0.4388  & 0.1631 & 0.1782  & 0.1388   \\
    222 & 0.5232  & 0.1839 & 0.1810  & 0.1334  \\
    400 & 0.1857  & 0.1007 & 0.1782  & 0.1388   \\
    420 & 0.3477  & 0.1406 & 0.1796  & 0.1362   \\
    422 & 0.4388  & 0.1631 & 0.1803  & 0.1347   \\
    \bottomrule
    \multicolumn{5}{l}{$^a$ The $c_{11}$, $c_{12}$, and $c_{44}$ values  (in GPa) for} \\
   \multicolumn{5}{l}{both compounds were taken from \cite{SS}} \\
    %\multicolumn{7}{l}{$^a$ $c_{11}=41.0$, $c_{12}=7.0$, $c_{44}=6.3$ GPa} \\
    %\multicolumn{7}{l}{$^b$ $c_{11}=49.7$, $c_{12}=13.0$, $c_{44}=12.8$ GPa} \\
    %\multicolumn{7}{l}{$^c$ $c_{11}=33.8$, $c_{12}=2.2$, $c_{44}=3.6$ GPa} \\
    \end{tabular}%
  \label{cs}%
\end{table}%

The crystal structure for alkali halides ---more commonly referred as rock-salts--- has been extensively studied and characterized. Even though their structure takes a cubic symmetry and either the well known FCC of BCC atomic packing, the bond's ionic nature impose a symmetry perturbation and thus extra deformation restrains. A burgers vector $b=\frac{1}{2}\av{100}$, for example, is not symmetrically allowed as a half shift would compromise the crystal's overall neutrality, and thus the main burgers vector is given by $b=\frac{1}{2}\av{110}$. Such restriction then translates, for example, into the edge dislocations being forbidden for the primary slip system for alkali halides $\av{110}\{110\}$, as the orthogonal relations between the slip direction, the burgers vector, and the dislocation axis system can not be simultaneously realized. In a similar manner, the screw dislocation is forbidden for the secondary $\av{100}\{110\}$ and tertiary $\av{111}\{110\}$ slip systems. While screw dislocations can have six different variants for the primary slip system, the orthogonal relationships between the burgers vector $(\hat{b})$, the dislocation line $(\hat{l})$ and the slip plane normal vector $(\hat{p})$ can not be simultaneously realized for edge dislocations. For example, in the case of the $(110)$ plane and the $[\bar{1}10]$ direction, $\hat{x_3}\parallel\hat{l}$ and $\hat{x_2}\parallel\hat{p}$. Because $\hat{b}\neq \av{001}$ in response to the crystal net charge neutrality, both  $\hat{b}$ and $\hat{x_1}\parallel\av{110}$, and thus the $\hat{x_1}\perp\hat{x_2}$ can not be fulfilled for that slip plane-direction combination. An analogous argument applies to each permutation of $\av{110}\{110\}$, and thus edge dislocations are geometrically forbidden in this system.  Overall, the primary slip system $\av{110}\{110\}$ allows only screw dislocations, while the secondary slip system $\av{100}\{110\}$ only permits edge dislocations to occur.
\begin{figure}[b!]   
% includes a figure in your pdf, the pictures should be in the image folder of your project.
	\includegraphics[width=0.45 \textwidth]{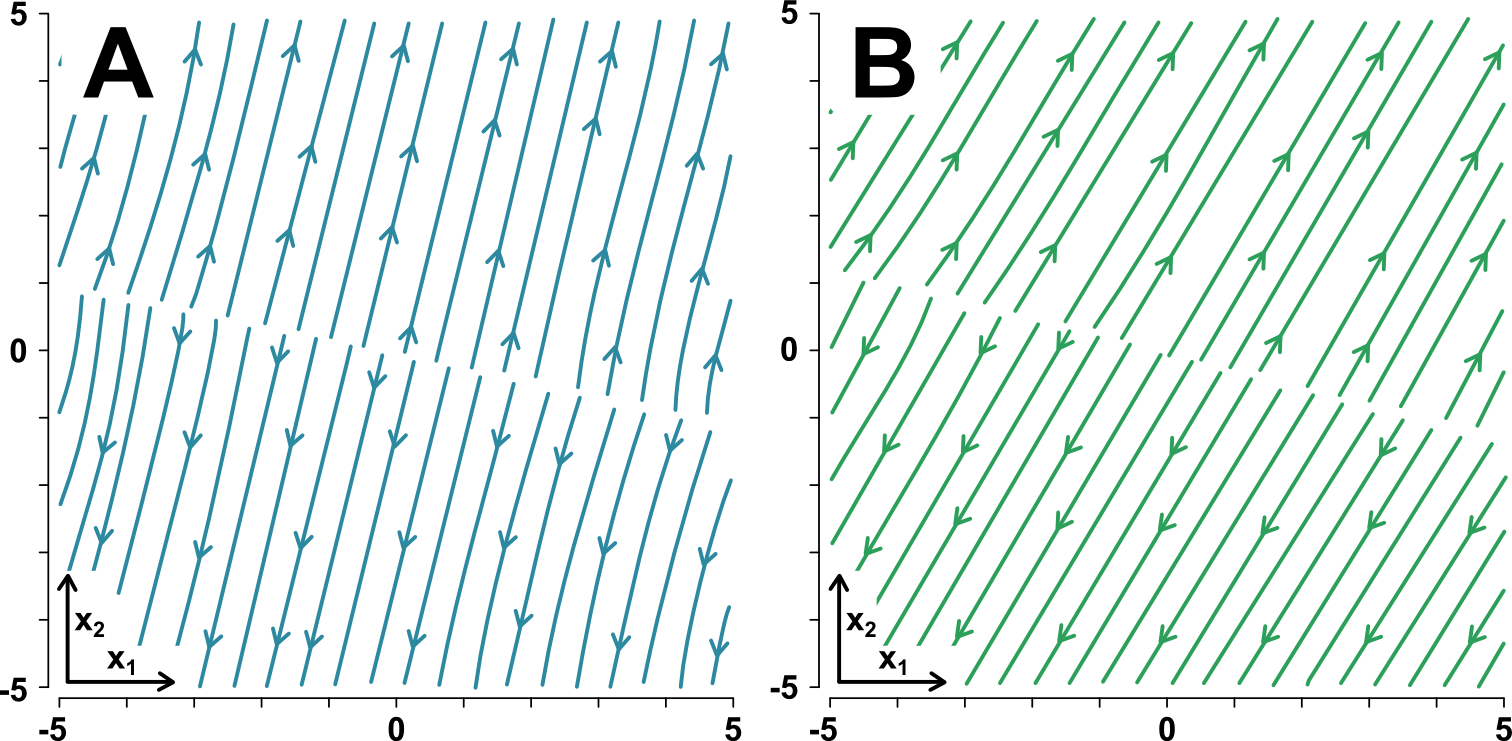}
	\caption{Graphical representations of the displacement field in the $\av{100}\{110\}$ edge slip system for (A) KCl and (B) NaCl.}
	\label{FFs}  % give your figure a lable to cite it with \ref{Fig:beach}
\end{figure}

The contrast factors for the allowed edge and screw slip-systems for both KCl and NaCl were computed with eqs. \eqref{C} by using the elastic constants of both ionic compounds as reported in \cite{SS}. These results are shown in Table \ref{cs}. The calculations for KCl suggest the $(111)$ as the most favorable plane for dislocation, closely followed by $(110)$. It is worth noticing that the obtained contrast factors for NaCl are significantly similar to each other, thus sugesting no overall preference for one plane against the other and that the dispersion in the integral breaths (i.e. divergence from theory) might be related to factors beyond strain anisotropy theory, such as those derived from plane faulting or twinning effects\cite{10}. The displacement fields for both KCl and NaCl powder samples are presented in Fig. \ref{FFs}. In particular, the $\av{100}\{110\}$ edge slip system on a dislocation coordinate system in which the $x_3$-axis is parallel to the dislocation line is showed. The plots exemplify the way on which the sliding lattice blocks move in opposite directions, with the lower and upper halves moving in and out of the page, respectively. As observed the KCl field exhibits a slight curvature on the edges, in agreement with the contrast factors calculated which showed a slight anisotropy. The NaCl field lines are overall mostly parallel and all are practically identical, in accordance with the contrast factors being nearly identical for different $(hkl)$ reflections.

\begin{figure}[t!]   
% includes a figure in your pdf, the pictures should be in the image folder of your project.
	\includegraphics[width=0.45 \textwidth]{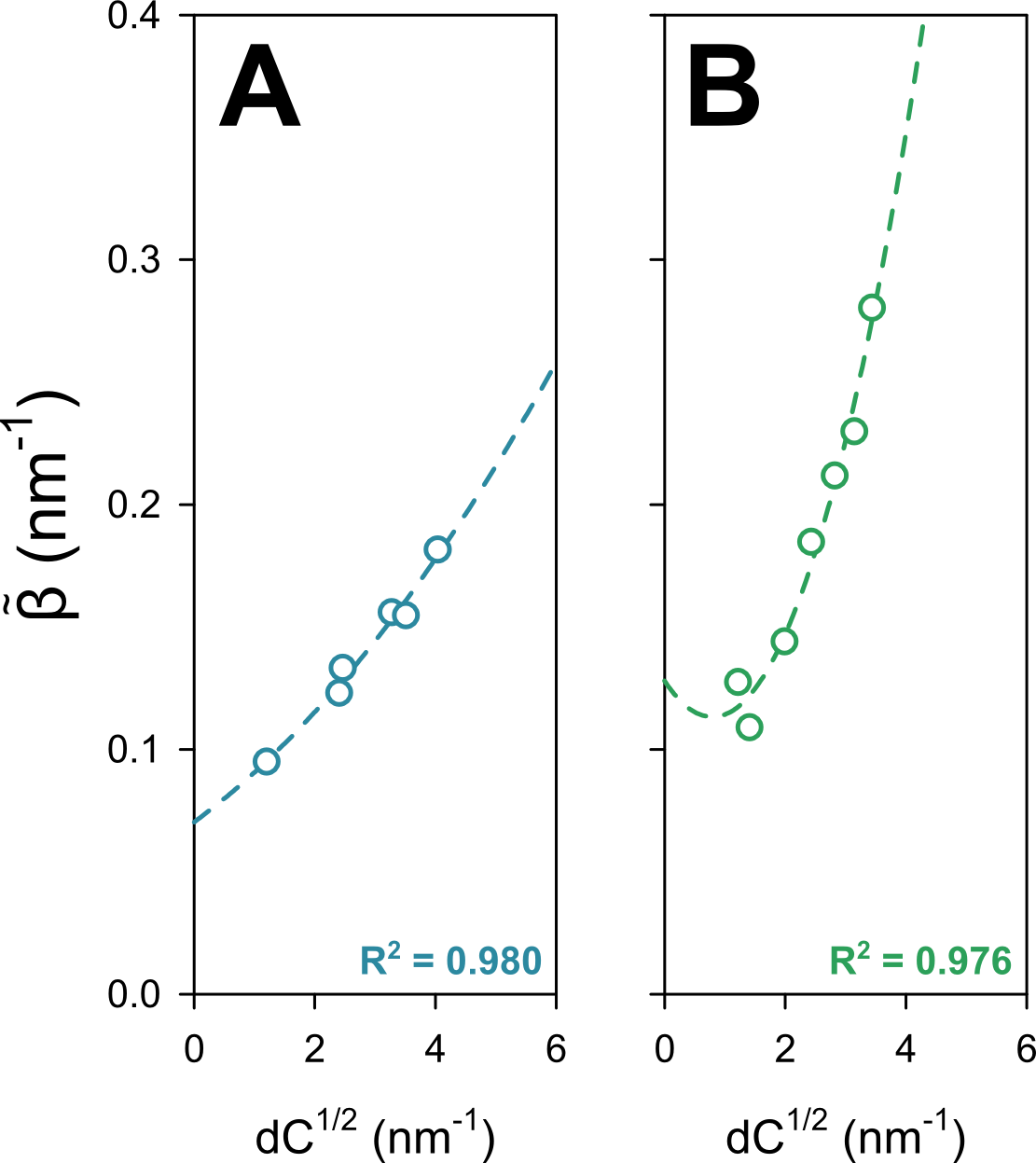}
	\caption{Modified WH plots in reciprocal space for powder samples of (A) KCl and (B) NaCl.}
	\label{MWHp}  % give your figure a lable to cite it with \ref{Fig:beach}
\end{figure}

Then, the calculated contrast factors were used to introduce a proper scaling factor for the reciprocal plane spacing and perform a modified WH analysis for both salts. The best data fits were obtained when the overall average contrast factors were calculated by assuming an equal proportion of screw/edge dislocations, thus suggesting that there is no preference for one dislocation type over the other. Both plots are presented in Fig. \ref{MWHp} and the corresponding fitting parameters are shown in Table \ref{qq}.Both samples showed a decrease in the apparent crystallite size when comparing the modified WH parameters to their classical counterparts; surprisingly, the modified apparent strain for NaCl is compressive, while a regular WH analysis would describe a tensile strain.  

\begin{table}[b!]
\setlength{\tabcolsep}{3pt}
\renewcommand{\arraystretch}{1.3}

\centering
  \caption{Numerical fittings for both WH and Modified WH analysis of samples}
    \begin{tabular}{cccccc}
    \toprule
     &\multicolumn{2}{c}{WH} &  \multicolumn{3}{c}{Modified WH}\\
     
    & $\tau$(nm) &  $\zeta$ (\%) & $\tau'\,$(nm) & $\zeta'\,$(\%) & $\tilde{\beta_2}$(nm)  \\
    \midrule
VO  & 13.68 &  1.18  & - & - & - \\ 
KCl & 17.86 & 0.74  & 14.22 & 0.90 & 2.2$\times10^{-3}$ \\ 
NaCl & 52.08 &  1.37  & 7.80 & -1.85 & 2.2$\times10^{-2}$ \\ 
    \bottomrule
    \end{tabular}%
  \label{qq}%
\end{table}%

On the one hand, the KCl fitting seems to be significantly improved from that of the classical analysis, with a relatively low $\tilde{\beta_2}$ coefficient (i.e. almost a linear behavior), corresponding to an anisotropic strain with a low fluctuation in the dislocation density. The NaCl fitting, on the other hand,  reveals a more pronounced parabolic behavior with a $\tilde{\beta_2}$ coefficient being almost ten-fold as that of KCl, thus suggesting a higher degree of correlation between the adjacent lattice points. The results obtained for KCl can be explained by the fact that the ionic bonds give rise to a lattice composed mainly by Ar cores on a large scale. This provides a more symmetric electronic environment, and thus a low degree of correlation is observed. Conversely, the NaCl lattice is composed of a mixture of Ne and Ar cores, and this pair coupling leads to a higher correlation degree.  

\section*{Conclusions}

The distortion tensor described in a stretched coordinate system allows for the straightforward computation of contrast factors for cubic slip-systems. Furthermore, the inclusion of symmetry constrains provides significant insights in the understanding of strain contributions. Within this context, this study extends the calculation of contrast factors to binary Fm$\bar{3}m$ materials. For the first time, the contrast factors for KCl and NaCl were calculated and used to characterize their microstructure through the modified WH analysis. 

%We believe that this approach can be further extended to other materials with challenging symmetries. 

\section*{Acknowledgments}
Financial support from the National Council of Science and Technology of Mexico (CONACyT, grant 269399) is greatly appreciated. 

\section*{APPENDIXES}
\setcounter{equation}{0}
\renewcommand{\theequation}{A\arabic{equation}}

%\setcounter{equation}{0}
\begin{comment}
\subsection*{A1. Geometric contribution in Contrast Factors}

As further described in \cite{11A}, each dislocation defines a new coordinate system on which $\hat{k}$ is defined to be parallel to $\hat{l}$, the dislocation line. The slip coordinate system for screw $\left(\hat{l}\parallel\hat{b}\right)$ and edge $\left(\hat{l}\perp\hat{b}\right)$ dislocation is defined as:
\beq
\begin{cases}
\hat{i} = \hat{j} \times \hat{k} \\
\hat{j} \perp \hat{p} \\
\hat{k} \parallel \hat{l}

\end{cases}
\eeq 

A transformation matrix between the crystallographic coordinate system and the dislocation one is then defined given by:

\begin{equation}
\textbf{P} = \begin{bmatrix}
\hat{i}\\\hat{j}\\\hat{k}
\end{bmatrix}
\end{equation}

\noi and thus the normalized diffraction vector $\bm{\hat{g}}$ can be transformed into $\bm{g'}$:
\begin{equation}
\bm{g'} = \bm{P}\,\bm{\hat{g}}
\end{equation}

\noi the components of $\bm{g'}$ in turn define the direction cosines $\left\{\gamma_i, i =1,2,3\right\}$
\end{comment}

\subsection*{A1. Slip Systems}

A \textit{slip system} is defined by a sliding plane $(hkl)$ which glides along or perpendicularly to a slip direction $[hkl]$. Each slip system can have a variety of realizations as defined by permutations of the plane and direction family components.  For each material, both its geometry and atomic packing determine what particular combinations of planes and directions are most likely to undergo deformations. A primary slip system, for example, is that on which the coupled plane and direction carry  simultaneously the largest planar and linear density, thus being the most likely to undergo deformations. A particular set of stress conditions, however, might lead to the expression of secondary or tertiary slip system for a particular deformation. Some geometries, however, even forbid some slip-system couplings to occur in the first place. 

\subsection*{A2. Distortion Displacement}

By expressing the relations in eq.\eqref{H} in either the polar coordinates $(r, \varphi)$ or the 'stretched' polar coordinates $(\rho,\phi)$, it can be easily shown that $\xi,\eta \propto r,\rho$ and $\sff{r} \propto r^{-1},\rho^{-1}$. Thus, the overall $r$ or $\rho$ dependence for the distortion displacement vanishes, and each component can be explicitly written as a function of $\varphi$ as given by:

\begin{eqnarray}
H_{1,R}(\varphi)  &=&  \phantom{-}\myfrac[2pt]{\cos(\varphi)+a\sin(\varphi)}{(\cos(\varphi)+a\sin(\varphi))^2 + c^2\sin^2(\varphi)} \nonumber \\
H_{1,I}(\varphi)  &=&   -\myfrac[2pt]{c\sin(\varphi)}{(\cos(\varphi)+a\sin(\varphi))^2 + c^2\sin^2(\varphi)} \nonumber \\
H_{2,R}(\varphi)  &=&  \phantom{-}\myfrac[2pt]{a\cos(\varphi)+|p|^2\sin(\varphi)}{(\cos(\varphi)+a\sin(\varphi))^2 + c^2\sin^2(\varphi)} \nonumber \\
H_{2,I}(\varphi)  &=&  \phantom{-}\myfrac[2pt]{c\cos(\varphi)}{(\cos(\varphi)+a\sin(\varphi))^2 + c^2\sin^2(\varphi)} \nonumber \\
\nonumber \\
\end{eqnarray}

In particular, the relationship between $\varphi$ and $\phi$, as derived by using the definitions in eq. \eqref{stretched} is given by:
\begin{equation}
\varphi = \arctan \left(\frac{\sin(\phi)}{c\cos(\phi)-a\sin(\phi)}\right)
\end{equation}

Thus, the distortion displacement can also be expressed as a function of the 'stretched' polar angle $\phi$ in a more symmetric way as given by:
\begin{eqnarray}
H_{1,R}(\phi)  &=&  \phantom{a\cos(\phi)+\,-}\,\cos(\phi) \cdot \Gamma(\phi)  \nonumber \\
H_{1,I}(\phi)  &=&   \phantom{a\cos(\phi)\,\,+} \;-\sin(\phi)\cdot \Gamma(\phi) \nonumber \\
H_{2,R}(\phi)  &=&  (a\cos(\phi)+c\sin(\phi))\cdot \Gamma(\phi) \nonumber \\
H_{2,I}(\phi)  &=&  (c\cos(\phi)-a\sin(\phi))\cdot \Gamma(\phi) \nonumber \\
\nonumber \\
\end{eqnarray}

\noi with $\Gamma(\phi)=\sqrt{(\cos(\phi)-\frac{a}{c}\sin(\phi))^2+\frac{1}{c^2}\sin^2(\phi)}$. 

\bibliography{main}

\newpage

\end{document}